\documentclass[pdflatex,sn-apa]{sn-jnl}% APA Reference Style
\usepackage{graphicx}%
\usepackage{multirow}%
\usepackage{amsmath,amssymb,amsfonts}%
\usepackage{amsthm}%
\usepackage{mathrsfs}%
\usepackage[title]{appendix}%
\usepackage{xcolor}%
\usepackage{textcomp}%
\usepackage{manyfoot}%
\usepackage{booktabs}%
\usepackage{algorithm}%
\usepackage{algorithmicx}%
\usepackage{algpseudocode}%
\usepackage{listings}%
\usepackage{natbib}%
\theoremstyle{thmstyleone}%
\theoremstyle{thmstyletwo}%
\newtheorem{remark}{Remark}%

\theoremstyle{thmstylethree}%

\let\code=\texttt

\newcommand{\pkg}[1]{{\fontseries{m}\fontseries{b}\selectfont #1}}
\usepackage{bm}
\usepackage{xfrac} %% for nice fractions in text
\usepackage{amsmath}
\usepackage{amssymb}
\usepackage{amsfonts}
\usepackage{fancyvrb}
\newcounter{inlineenum}
\renewcommand{\theinlineenum}{\alph{inlineenum}}
\newenvironment{inlineenum}
{\unskip\ignorespaces\setcounter{inlineenum}{0}%
	\renewcommand{\item}{\refstepcounter{inlineenum}{\textit{\theinlineenum})~}}}
{\ignorespacesafterend}
\DefineVerbatimEnvironment{Code}{Verbatim}{}

\DeclareMathOperator{\Tr}{tr}

\DeclareMathOperator{\Var}{var}
\DeclareMathOperator{\rank}{rank}
\DeclareMathOperator{\pev}{pev}

\newcommand{\matr}[1]{\ensuremath{\bm{#1}}}     % ISO complying version
\newcommand{\tp}{{\!\scriptscriptstyle \top}}

\newcommand{\newAcrit}{\ensuremath{\mathcal{A}}-criterion}

\newcommand{\odw}{\pkg{odw}}

\newcommand{\Wone}{\ensuremath{\bm{W}_{1}}}
\newcommand{\Woneone}{\ensuremath{\bm{W}_{11}}}
\newcommand{\Wonetwo}{\ensuremath{\bm{W}_{12}}}
\newcommand{\Wtwo}{\ensuremath{\bm{W}_{2}}}

\newcommand{\Xoneone}{\ensuremath{\bm{X}_{11}}}
\newcommand{\Xonetwo}{\ensuremath{\bm{X}_{12}}}
\newcommand{\Xtwo}{\ensuremath{\bm{X}_{2}}}

\newcommand{\Zoneone}{\ensuremath{\bm{Z}_{11}}}
\newcommand{\Zonetwo}{\ensuremath{\bm{Z}_{12}}}
\newcommand{\Ztwo}{\ensuremath{\bm{Z}_{2}}}
\newcommand{\bone}{\ensuremath{\bm{\beta}_{1}}}
\newcommand{\boneone}{\ensuremath{\bm{\beta}_{11}}}
\newcommand{\bonetwo}{\ensuremath{\bm{\beta}_{12}}}
\newcommand{\btwo}{\ensuremath{\bm{\beta}_{2}}}

\newcommand{\toneone}{\ensuremath{\bm{\tau}_{11}}}
\newcommand{\tonetwo}{\ensuremath{\bm{\tau}_{12}}}
\newcommand{\ttwo}{\ensuremath{\bm{\tau}_{2}}}

\newcommand{\uoneone}{\ensuremath{\bm{u}_{11}}}
\newcommand{\uonetwo}{\ensuremath{\bm{u}_{12}}}
\newcommand{\utwo}{\ensuremath{\bm{u}_{2}}}
\newcommand{\Gone}{\ensuremath{\bm{G}_{1}}}
\newcommand{\Goneone}{\ensuremath{\bm{G}_{11}}}
\newcommand{\Gonetwo}{\ensuremath{\bm{G}_{12}}}
\newcommand{\Gtwo}{\ensuremath{\bm{G}_{2}}}
\newcommand{\Gones}{\ensuremath{\bm{G}^*_{1}}}
\newcommand{\Goneones}{\ensuremath{\bm{G}^*_{11}}}
\newcommand{\Gonetwos}{\ensuremath{\bm{G}^*_{12}}}
\newcommand{\Gtwos}{\ensuremath{\bm{G}^*_{2}}}

\newcommand{\soneone}{\ensuremath{\bm{\sigma}_{g_{11}}}}
\newcommand{\sonetwo}{\ensuremath{\bm{\sigma}_{g_{12}}}}
\newcommand{\stwo}{\ensuremath{\bm{\sigma}_{g_2}}}
\newcommand{\Mtwo}{\ensuremath{\bm{M}_{2}}}
\newcommand{\tbone}{\ensuremath{\tilde{\bm{\beta}}_{1}}}
\newcommand{\tbtwo}{\ensuremath{\tilde{\bm{\beta}}_{2}}}

\newcommand{\Gmoneone}{\ensuremath{\bm{G}^{-1}_{11}}}
\newcommand{\Gmonetwo}{\ensuremath{\bm{G}^{-1}_{12}}}
\newcommand{\Gmtwo}{\ensuremath{\bm{G}^{-1}_{2}}}
\newcommand{\Coneone}{\ensuremath{\bm{C}_{1}}}
\newcommand{\Vtwo}{\ensuremath{\bm{V}_{2}}}
\newcommand{\Ptwo}{\ensuremath{\bm{P}_{2}}}

\usepackage{listings}
\usepackage{newfloat}
\DeclareFloatingEnvironment[fileext=lop,placement={htb},name=Listing]{lstfloat}
\usepackage{caption}
\usepackage{lmodern} %for bold ttfamily
\definecolor{greyColour}{rgb}{0.95,0.95,0.95}
\lstdefinelanguage{none}{identifierstyle=\color{black},morekeywords={odw, summary, table, update},keywordstyle=\color{black}\bfseries}
\begin{document}

\title{The design of selection experiments using a model-based approach}


\author*[1]{\fnm{Brian} \sur{Cullis}}\email{bcullis@uow.edu.au}

\author[1]{\fnm{Alison} \sur{Smith}}

\author[1]{\fnm{David} \sur{Hughes}}

\author[1]{\fnm{David} \sur{Butler}}

\affil[1]{\orgdiv{MMaED, School of Mathematics and Applied Statistics}, \orgname{University of Wollongong, Wollongong}, \orgaddress{\city{Wollongong}, \postcode{2522}, \state{NSW}, \country{Australia}}}


\abstract{Plant breeding programs use data obtained from multi-environment selection experiments to produce improved varieties with the ultimate aim of maintaining high levels of genetic gain. Selection accuracy can be improved with the use of advanced statistical analytical methods that use informative and parsimonious variance models for the set of genotype by environment interaction effects, include information on genetic relatedness and appropriately accommodate non-genetic sources of variation within the framework of a single step estimation and prediction algorithm. Maximal gains from using these advanced techniques are more likely to be achieved if the  designs used match the aims of the selection experiment and make full use of the available resources. In this paper we present an approach for constructing designs for selection experiments which are optimal or near optimal against a robust and sensible linear mixed model. This model reflects the models used for analysis. The approach is flexible and introduces an additional step to accommodate efficient resource allocation of replication status to genotypes, which is undertaken prior to the allocation of plots to genotypes. A motivating example is used to illustrate the approach, two illustrative examples are presented one each for single and multiple environment selection experiments and several in-silico simulation studies are used to demonstrate the advantages of these approaches.}

\keywords{optimal design, correlated treatment effects, linear mixed model,
	selection experiments, resource allocation, multi-environment selection experiments} %% First letter not capped


\maketitle

	\section{Introduction}\label{sec:intro} 
The aim of plant breeding programs is to produce new varieties with superior genetic value for commercially important traits such as disease resistance, grain yield and end-use quality. The process of plant breeding is a multi-year/multi-cycle collection of data \citep{ariefUtilizationMultiyearPlant2019} in which lines are created from a fixed number of crosses between elite parents. This phase is referred to as generation phase one (GP1). The first filial generation (F1s) undergo marker assisted selection for simple and highly heritable traits such as disease resistance. Superior material from GP1 are selfed (from second to fourth filial generation, F2 to F4) in GP2. During this phase, so-called test lines (i.e.~F4s) undergo further selection for traits are either influenced by a small number of genes or have low levels of genotype by environment interaction. These traits are associated with either biotic or abiotic factors such as: phytophthora root rot ({\it Phytophthora medicaginis}) in chickpeas (\cite{asifRapidHighThroughput2023}), chocolate spot (\emph{Botrytis fabae}) in faba beans, cold tolerance in rice, acid soil tolerance in lentils and ascochyta blight (\emph{Ascochyta rabei}) in chickpeas. Screening for these traits involves the design and analysis of a single environment selection experiment (SESE). These experiments usually utilise controlled environment facilities, so that large numbers of test lines can be assessed with reduced cost in a timely manner. There are often important constraints, such as seed supplies of test lines and capacity of infra structure. Hence these experiments are characterised by minimal and unequal replication of test lines (\cite{asifRapidHighThroughput2023}). More recently, plant breeding programs implement genomic selection (GS) \citep{goddardUsingGenomicRelationship2011} for other quantitative traits in these phases, however the extent which GS has been implemented depends on the prediction accuracy of GS \citep{meuwissenAccuracyGenomicSelection2012} for that trait.

In GP3 a reduced number of test lines which have been promoted from GP2 are extensively phenotyped for disease, yield and end-use quality. In most plant breeding programs GP3 comprise four stages of testing, where each stage represents one year. These stages will be referred to as stage one (S1), stage two (S2), stage three (S3) and stage four (S4). Annually for each stage, test lines, advanced breeding lines and check varieties are assessed in a designed experiment which consist of trials located at a pre-determined set of geographic locations. Such an experiment will be referred to as a multiple environment selection experiment (MESE). The locations in the MESE are chosen to represent growing regions of commercial significance. In this paper, ``genotype'' is used to describe the set of test lines (TLs), advanced breeding lines (ABLs) and check varieties (ChVs). A TL is a genotype which has been tested in S1 or S2 only, an ABL is a genotype which has been tested in S1 and promoted to S2 for at least two years and/or promoted to S3 and S4. A commercial variety is referred to as a  ChV. The term environment is synonymous with site and represents a unique combination of geographic location and year of planting. The term ``trial'' refers to the set of plots at each location within a MESE. It often happens that there may be several trials at the same environment which are components of different MESEs (i.e.~from different stages of testing). These trials which are at the same location within the same year but part of different MESEs are referred to as co-located trials. GP4 involves evaluation and testing of superior ABLs for uniqueness (Plant Breeding Rights), quality classification and independent evaluation for yield performance. This testing phase is outside the focus of this paper. 

The total genetic value for overall performance and stability of genotypes is obtained from the analysis of a collection of MESEs. This collection must span both a sufficient number of location and years to achieve the desired accuracy of prediction. Accordingly, data-sets can be constructed for such analyses using the ideas of \cite{smithUseContemporaryGroups2021,lisleInformationBasedDiagnostic2021}, while the preferred method of analysis is a single-step FALMM \cite{verbylaTwostageAnalysisMultienvironment2023,smithPlantVarietySelection2021}. 

To motivate the work presented in this paper, a set of MESEs is described in the following section. These MESEs were conducted in 2023 by an Australian wheat breeding program and were designed using the methods described in this paper. These MESEs highlight some of the fundamental requirements, constraints and aims of a MESE. The example also provides compelling evidence of the need to incorporate genetic relatedness in the construction of optimal or near-optimal designs of selection experiments. 

\subsection{Motivating Example}\label{ssec:motex} 

The motivating examples are provided by Intergrain and involve the MESEs conducted in 2023 by the hard wheat breeding program and the MESE for S2 conducted in 2022 by the barley breeding program. Here we describe the MESEs for the hard wheat breeding program, while the S2 MESE for the barley breeding program will be described and used as the illustrative example in part two of the series. 

The aim of each MESE is to provide data for inclusion in a collection of MESEs spanning 2019 to 2023. This collection of MESEs is then used as input data for analysis which provides information for selection decisions in 2023/2024. These selection decisions are \begin{inlineenum}
	\item promotion of TLs in S1 in 2023 to S2 in 2024;
	\item promotion of TLs and ABLs in S2 in 2023 to S3 in 2024;
	\item promotion of ABLs in S3 in 2023 to S4 in 2024;
	\item promotion of ABLs in S4 in 2023 to GP4
\end{inlineenum}. Tables~\ref{tab:esump} and~\ref{tab:esums} present comprehensive summaries of the MESE for S1, S2, S3 and S4 respectively.

The MESE for S1 consisted of 3954 TLs and 22 ChVs which were grown at 11 locations. Each trial involved two management blocks and each comprised 564 plots laid out in a rectangular lattice with 12 columns and 47 rows. The two management blocks were row-adjacent resulting in a $12\times 94$ rectangular lattice spanning both blocks within the trial. The major limiting factor for S1 MESEs is seed supply for the TLs. In this MESE, TLs had sufficient seed to be tested in a minimum of one plot to a maximum of five plots in the full experiment. Given the substantial genotype by environment interaction for yield and end-use quality \citep{heslotIntegratingEnvironmentalCovariates2014a, fairlieGenotypebyenvironmentInteractionWheat2024} it is logical that each TL should be tested in as many locations. However, it is also desirable that some genotypes are replicated within a trial so that there is sufficient information for estimation of variance parameters associated with non-genetic effects and error associated with each trial. In this example, the strategy was to use additional plots of ChVs in each trial. The median number of plots per ChV per trial was 5.07 and this resulted in a mean replication frequency of genotypes within a trial of 5\%. The low level of replication of TLs results in low infill for the genotype by environment concurrence matrix (GECM). The ``infill'' of a matrix is the number of non-zero elements divided by the product of the dimensions of the matrix. The infill of the GECM for TLs is very low (0.37). The infill of the GECM for ChVs is 0.61, but this is somewhat misleading as there is a core set of 10/24 ChVs which were tested in all trials, while the remaining 14 ChVs were tested in trials which were at locations of relevance to each ChV respectively.

The primary aim of the S2 MESE is to test the subset of TLs which have been promoted from S1 in 2022. It is often also desirable to re-test a subset of ABLs which were in the S2 MESE in 2022. As for the S1 MESE, breeders require that both TLs and ABLs be tested at as many locations as possible subject to seed and budgetary constraints. In general, there is more seed available for each TL and ABL, but the distribution of seed resources needs to consider both sufficient sampling of environments but maintain an adequate replication of genotypes within trials. There were a total of 25 locations (i.e.~trials) in the S2 MESE  and each trial consisted of 552 plots arranged in a $12\times 46$ rectangular lattice (see Table~\ref{tab:esump}). The minimum, median and maximum number of plots in the full S2 MESE were $(5,13,15)$ and $(9,13,50)$ for TLs and ABLs respectively. The minimum, median and maximum number of trials which tested each TL and ABL were $(3,9,11)$ and $(7,9,25)$ for TLs and ABLs respectively (see Table~\ref{tab:esums}). The infill of the GECMs for TLs and ABLs are 0.37 and 0.38 respectively. As for the S1 MESE, there was a core set of 7/12 ChVs which were grown at all sites. The mean replication frequency of genotypes within a trial was 40\%.

In the S3 and S4 MESEs, a reduced set of ABLs are tested at more locations, with increased replication of genotypes within trials. The data in Tables~\ref{tab:esump} and~\ref{tab:esums} illustrates that the infill of the GECM for ABLs is near 1 for both MESEs, with an approximate replication frequency of ABLs of two plots in each trial.  

Pedigree information was available for 5738 genotypes which comprised 5154 genotypes which were tested in the MESEs conducted in 2023, and an additional 616 ancestral genotypes which were not deployed in the MESEs. Marker information was also available 4306 genotypes from these MESEs.

Although the primary aim of the MESEs is to provide data for selection decisions on the set of TLs and ABLs, given the emerging importance of GS for identifying test lines with superior additive genetic values for yield and end-quality in GP2, plant breeding programs need to develop cost-effective and sensible training populations for these traits. 
\cite{meuwissenAccuracyGenomicSelection2012} identified and quantified the factors which underlie and determine the accuracy of GS.  These are
\begin{inlineenum}
	\item marker density;
	\item size of the training population;
	\item trait heritability;
	\item genome size and structure;
	\item historical effective population size;
	\item relationship between training population and selection candidates;
	\item the number of genes and distribution of the effects;
	\item methods used for the estimation of marker effects
\end{inlineenum}. The continued advancement of high throughput genotyping technologies have led to the widespread availability of dense panels of molecular markers (for example SNPs). Key traits such as yield and end-use quality are often characterised by a large number of small effect loci \citep{schmidtPredictionMaltingQuality2016}, have low heritability and exhibit high levels of genotype by environment interaction \citep{heslotIntegratingEnvironmentalCovariates2014a, bhattaMultitraitGenomicPrediction2020, jarquinReactionNormModel2014,fairlieGenotypebyenvironmentInteractionWheat2024}.

Based on these findings, it is obvious that so-called in-situ training populations based on a collection of SESEs and/or MESEs spanning an adequate set of years and locations would produce the most effective and desirable training population for a breeding program. Hence, construction of the design for MESEs must also consider this objective so that higher levels of genetic gain are maintained. 

Furthermore, it is clear that simple or piecemeal approaches to design construction for SESEs or MESEs in S1 and S2 would be unlikely to produce an optimal or near optimal design. For example,  choosing the subset of genotypes to replicate should consider genetic relatedness between the genotypes, and this choice must be subject to seed and experimental resources. Similarly, allocation of genotypes to sites, subject to constraints, should consider genetic relatedness between genotypes. The current practice for S3 and S4 MESEs is to use standard designs with fixed genotype effects for each environment. Given the aim is selection and interest lies in prediction of overall performance and stability \citep{piephoMethodsComparingYield} across all or a subset of environments within the MESE, then this approach would seem sub-optimal. The trade off between infill of the GECM versus replication within each environment within the MESE is also a complex problem. Lastly, the additional aim of SESEs or MESEs is to be part of a suitable training population for GS calibration, hence the design construction process must include genetic relatedness to produce efficient designs for additive effects (see later in this section). 

% latex table generated in R 4.4.1 by xtable 1.8-4 package
% Wed Nov  6 13:31:28 2024
\begin{table}[ht]
	\centering
	\scalebox{1}{
		\begin{tabular}{ccccc}
			\toprule   Attribute & S1 & S2 & S3 & S4\\
			\midrule Plots (P) & 12408 & 13800 & 4608 & 5400 \\ 
			Sites (S) & 11 & 25 & 24 & 33 \\ 
			Blocks(B) in S & 2 & 1 & 1 & 1 \\ 
			P in B in S & 564 & 552 & 192 & 168(30), 120(3) \\ 
			Col $\times$ Row & $12\times 47$ & $12\times 46$ & $12\times 16$ & $12\times 14$ (30), $6\times 20$, (3) \\ 
			TL & 3954 & 911 & 0 & 0 \\ 
			ABL & 0 & 119 & 92 & 43 \\ 
			ChV & 22 & 12 & 14 & 33 \\ 
			infill(TL) & 0.27 & 0.37 &  &  \\ 
			infill(ABL) &  & 0.38 & 0.99 & 0.96 \\ 
			infill(ChV) & 0.61 & 0.71 & 0.38$^*$ & 0.62 \\ 
			\bottomrule \end{tabular}
	}
	\caption{Primary environment attributes for the motivating example. 
		$^*$: There is a base set of 5 out of 14 check varieties grown at all sites. Infill represents the proportion of non-zero cells in the cross
		tabulation with site. Values in brackets for the column and row labelled S4 and Col$\times$ Row respectively refer to the 
		number of trials for that array size.} 
	\label{tab:esump}
\end{table}

% latex table generated in R 4.4.1 by xtable 1.8-4 package
% Wed Jan 15 19:39:06 2025
\begin{table}[ht]
	\centering
	\scalebox{1}{
		\begin{tabular}{ccccccccccccc}
			\toprule    &\multicolumn{3}{c}{S1} & \multicolumn{3}{c}{S2} &
			\multicolumn{3}{c}{S3} & \multicolumn{3}{c}{S4} \\
			Attribute & Min & Med & Max & Min & Med & Max & Min & Med & Max & 
			Min & Med & Max \\
			\midrule TL in S & 1054 & 1057 & 1060 & 322 & 337 & 351 &  &  &  &  &  &  \\ 
			ABL in S &  &  &  & 36 & 47 & 53 & 91 & 91 & 92 & 40 & 41 & 43 \\ 
			ChV in S & 12 & 14 & 15 & 8 & 8 & 10 & 5 & 5 & 14 & 19 & 21 & 22 \\ 
			S in TL & 1 & 3 & 5 & 3 & 9 & 11 &  &  &  &  &  &  \\ 
			S in ABL &  &  &  & 7 & 9 & 25 & 1 & 24 & 24 & 13 & 33 & 33 \\ 
			S in ChV & 1 & 7 & 11 & 1 & 25 & 25 & 1 & 1$^*$ & 24 & 5 & 23 & 33 \\ 
			P in TL & 1 & 3 & 5 & 5 & 13 & 15 &  &  &  &  &  &  \\ 
			P in ABL &  &  &  & 9 & 13 & 50 & 1$\dag$ & 48 & 48 & 26 & 96$\ddag$ & 96 \\ 
			P in ChV & 1 & 30 & 71 & 1 & 50 & 51 & 1 & 1$^*$ & 49 & 10 & 46 & 96 \\ 
			P in S in TL & 1.00 & 1.00 & 1.00 & 1.37 & 1.39 & 1.43 &  &  &  &  &  &  \\ 
			P in S in ABL &  &  &  & 1.31 & 1.43 & 1.50 & 1.86 & 2 & 2.00 & 2.00 & 2.95 & 2.98 \\ 
			P in S in ChV & 4.60 & 5.07 & 5.92 & 1.88 & 2.00 & 2.11 & 1.5 & 2 & 2.00 & 2.00 & 2.23 & 2.26 \\ 
			P in S in Genotype & 1.05 & 1.05 & 1.06 & 1.39 & 1.40 & 1.44 & 1.81 & 2 & 2.00 & 2.00 & 2.71 & 2.71 \\ 
			\bottomrule \end{tabular}
	}
	\caption{Secondary environment attributes table for the motivating example. 
		$^*$: There is a base set of 5 out of 14 check varieties grown at all sites. $\dag$: 91/92 have  
		48 plots. $\ddag$: 89/92 have 96 plots. } 
	\label{tab:esums}
\end{table}

\subsection{Background}\label{ssec:back}
There is a long and rich history concerned with the design of experiments involving a large number of treatments. The classical approach uses designs from the class of complete or incomplete block designs (see for example \cite{johnCyclicComputerGenerated1995,baileyDesignComparativeExperiments2008}). These designs hae been widely used in plant breeding programs for many years, however recently there has been an increasing interest in constructing and using designs which are model-based. Model-based designs are flexible and have many advantages over classical designs. Classical designs can also be easily constructed using a model-based approach \citep{butlerOptimalDesignExperiments2013,coombesDiGGeRSpatialDesign2009,coombesReactiveTABUSearch2002}, but importantly model-based designs can be linked to the model, or a model which approximates the model, used for the subsequent analysis of data obtained from the experiment. Examples of optimal or near-optimal model-based designs which cannot be constructed using classical approaches include designs for experiments with dependent errors \citep{butlerDesignFieldExperiments2014,martinEfficientBlockDesigns1991,martinConstructionOptimalNearoptimal1997,martinEfficientExperimentalDesigns2006}. Similarly the construction of optimal $p$-rep designs after \cite{cullisDesignEarlyGeneration2006} requires a model-based approach. A $p$-rep design is an alternative to grid plot designs \citep{kemptonDesignAnalysisUnreplicated1982} for use in early generation variety trials, and these designs are widely used in plant breeding programs. 

A major deficiency of classical designs and many model-based designs is that they do not address the aim of a selection experiment. If treatment effects are assumed fixed, then this does not match the method of analysis, but more importantly does not lead to optimal prediction of treatment effects using best linear unbiased prediction \citep{robinsonThatBLUPGood1991}. Additionally, assuming fixed treatment effects in design construction precludes the use of (genetic) relatedness. More recently \cite{vo-thanhGeneratingDesignsComparative2023} used a new search algorithm to generate augmented row-column designs. Their model was a linear (fixed-effects) model, where treatment, row and column effects were assumed to be fixed. \cite{cullisDesignEarlyGeneration2006} considered random treatment effects on the premise that this is a more sensible assumption when the aim is selection. They raised the possibility of using genetic relatedness, however they did not pursue this citing that most analyses of plant improvement data at that time, did not utilise genetic relatedness. It is now common place to use either ancestral or marker based forms of genetic relatedness in the analysis of data-sets arising from selection experiments in plant breeding programs and genomic studies (see for example,  \cite{oakeyJointModellingAdditive2007,oakeyJointModelingAdditive2006,oakeyGenomicSelectionMultienvironment2016,normanIncreasedGenomicPrediction2017,tolhurstGenomicSelectionMultienvironment2019}). \cite{buenofilhoBlockDesignsRandom2007} considered model-based design with correlated genetic effects, proposing a simple linear mixed model which included additive genetic effects and a single blocking factor. \cite{piephoComparisonExperimentalDesigns2006} examined three types of design for a simple genetic structure and concluded that designs which included the genetic structure associated with the treatment effects were superior to those designs which were constructed using fixed treatment effects. 

\cite{butlerDesignFieldExperiments2014} considered the design of plant breeding experiments which may have a complex plot structure and dependent errors, where the treatment effects are correlated. They illustrated their approach using a model-based paradigm for design construction using three examples taken from plant breeding with inbred and non-inbred species. Following on from this, \cite{cullisDesignEarlyStagePlant2020} proposed a model-based approach for generating optimal or near-optimal designs for SESEs. Their approach was similar to \cite{butlerDesignFieldExperiments2014} but they used informed variance parameter values in the design search. They also presented a novel updating scheme for calculation of optimality criteria and showed that this significantly reduced computing time per tabu \citep{gloverTabuSearchPart1989} loop. They conducted a comprehensive in silico study that compared three design types across a range of ancillary treatment factors and showed the advantage of a model-based design which used genetic relatedness compared with two other designs which did not use genetic relatedness, in terms of accuracy for the prediction of total genetic effects.  

The aim of this paper is to develop a model-based approach which extends the ideas of \cite{cullisDesignEarlyStagePlant2020,butlerDesignFieldExperiments2014}. An additional step is introduced which considers the allocation of replication to genotypes in SESEs or MESEs. While $p$-rep designs are widely used for the design of selection experiments in plant breeding and \cite{cullisDesignEarlyStagePlant2020} has recently demonstrated the importance of using genetic relatedness to allocate plots to genotypes. These approaches have not considered how best to choose the subset of genotypes to replicate. For example, in a simple $p$-rep design, this would involve choosing the subset of TLs to have two replicates rather than one. 
%
% this paragraph needs a rewrite
%
The paper is arranged as follows. In ??Material and Methods?? the general linear mixed model for model-based design with the inclusion of genetic relatedness is described, with particular attention given to the application for design construction of a SESE. Key elements of model-based design are presented and the search algorithm and computational elements used in the {\sf R} package  {\odw} \citep{butlerOptimalDesignLinear2018} are summarised. The stages in the design algorithm for construction of an optimal or near-optimal design for an SESE are then described in detail. In ''Results and Discussion'' these ideas are used to construct an efficient design for a SESE disease screening nursery in chickpeas and the results of an in silico simulation experiment designed to assess benefits of using new algorithm for the design of a $p$-rep design are presented and discussed.

\section{Materials and methods}\label{sec:stats}
Model-based design searches the design space for a configuration which is optimal in some sense under a pre-specified (linear) model. The model is usually chosen to be as close as possible to that expected for analysis. The design space is searched by an optimisation strategy which aims to find the configuration which minimises a suitably chosen optimality criteria. To sensibly describe the design construction algorithm for SESEs a brief overview of model-based design is presented. 

\subsection{Principles and concepts model-based designs: A brief overview}\label{ssec:mbds}

\subsubsection{The linear mixed model}\label{sssec:lmmdes}
Consider the construction of designs under the following model for the $n-$vector $\matr{y}$
\begin{equation}
	\matr{y} = \matr{X} \matr{\tau} + \matr{Z} \matr{u} + \matr{e}
	\label{eq:lmm}
\end{equation}
where $\matr{\tau}$ is an $r-$vector of fixed effects, $\matr{u}$ is a $z-$vector of random effects, $\matr{e}$ is the $n-$vector of errors and $\matr{X}$ and $\matr{Z}$ are the associated design matrices (for $\matr{\tau}$ and $\matr{u}$). So that Eqn~(\ref{eq:lmm}) aligns with the computing algorithm used in the {\sf R} package {\odw} \citep{butlerOptimalDesignLinear2018}, which is used  in this paper to construct optimal or near-optimal designs, the elements of $\matr{\beta} = [\matr{\tau}^\tp \; \matr{u}^\tp]^\tp$ are reordered so that Eqn~(\ref{eq:lmm}) is given by
\begin{equation}
	\matr{y} = \matr{W} \matr{\beta} + \matr{e}
	\label{eq:lmmw}
\end{equation}
with $\matr{W} = [\Wone \; \Wtwo]$, $\Wone = [\Woneone \; \Wonetwo]$, $\Woneone = [\Xoneone \; \Zoneone]$, $\Wonetwo = [\Xonetwo \; \Zonetwo]$ and $\Wtwo = [\Xtwo \; \Ztwo]$. Where possible, dimensions of the design sub-matrices and associated vectors of effects are omitted, as this simplifies the pedagogical presentation of the models and mathematical results.     

The matrix $\Wone$ is the design matrix for the set of \code{Permute} effects, $\bone = [\boneone^\tp \; \bonetwo^\tp]^\tp$. The \code{Permute} set, $P$, is impacted by the exchange policy used during the design search. The matrix $\Wtwo$ is the design matrix for the set of \code{Static} effects $\btwo = [\ttwo^\tp \; \utwo^\tp]^\tp$. This set is denoted by $S$ and this set is not impacted by the exchange policy. It typically includes effects associated with the plot structure of the experiment, or those associated with extraneous covariates or factors. Note $P\cap S = \varnothing$. 

The set $P$ is partitioned into the \code{Objective} set ($O$) and the \code{Linked} set ($L$). The matrices $\Woneone$ and $\Wonetwo$ are the design matrices associated with the vectors $\boneone$ and $\bonetwo$ of effects in $P$ and $L$ respectively. Lastly, to cater for complete generality, the vectors $\boneone = [\toneone^\tp \; \uoneone^\tp]^\tp$ and $\bonetwo = [\tonetwo^\tp \; \uonetwo^\tp]^\tp$, so that the sets $O$ and $L$ can contain both fixed and random effects respectively. It follows that $O\subseteq P$, $L\subseteq P$, $O \cup L = P$ and $O\cap L = \varnothing$. The set of effects in $O$ contribute to calculation of the optimality criteria, while the set of effects in $L$ do not contribute to calculation of the optimality criteria.	

\begin{remark}
	If a compound model term is in $P$ then any model term which is marginal to that term, other than the overall mean, cannot be in $S$.
\end{remark}
For example if \code{A:B} is in $P$ then neither \code{A} nor \code{B} can be in $S$. Use of such a LMM in {\odw} with this property would produce a flawed design. 

The random effects $\matr{u}$ and errors $\matr{e}$ in Eqn~(\ref{eq:lmm}) are assumed normally distributed such that
\begin{displaymath}
	\begin{bmatrix}\uoneone \\ \uonetwo \\ \utwo \\ \matr{e}  \end{bmatrix}
	\sim \text{N} \left(
	\begin{bmatrix}
		\matr{0} \\ \matr{0} \\ \matr{0} \\ \matr{0}
	\end{bmatrix},
	\begin{bmatrix}
		\Goneone & \matr{0} & \matr{0} & \matr{0} \\
		\matr{0} & \Gonetwo & \matr{0}  & \matr{0} \\
		\matr{0} & \matr{0} & \Gtwo & \matr{0}\\
		\matr{0} & \matr{0} & \matr{0} & \matr{R}		
	\end{bmatrix}
	\right)
\end{displaymath}
where $\Goneone, \Gonetwo$, $\Gtwo$ and $\matr{R}$ are positive definite matrices assumed to be functions of vectors $\soneone, \sonetwo, \stwo$ and $\matr{\sigma}_r$ respectively. Model-based design requires values for these parameters.

\subsubsection{Mixed model equations for the set of permute effects}
\label{sssec:estimation}
The full set of mixed model equations (MME) \citep{robinsonThatBLUPGood1991} for Eqn~(\ref{eq:lmmw}) can be written succinctly as
\begin{equation}
	\matr{C}\tilde{\matr{\beta}} = \matr{W}^\tp\matr{R}^{-1}\matr{y}
	\label{eq:mmc}
\end{equation}
and in full by
\begin{equation}
	\begin{bmatrix}
		\Wone^\tp\matr{R}^{-1}\Wone + \Gone^* & \Wone^\tp\matr{R}^{-1}\Wtwo\\
		\Wtwo^\tp\matr{R}^{-1}\Wone & \Wtwo^\tp\matr{R}^{-1}\Wtwo + \Gtwo^{*}
	\end{bmatrix}
	\begin{bmatrix}
		\tbone \\
		\tbtwo
	\end{bmatrix}
	=
	\begin{bmatrix}
		\Wone^\tp\matr{R}^{-1}\matr{y}\\
		\Wtwo^\tp\matr{R}^{-1}\matr{y}
	\end{bmatrix}
	\label{eq:mm} 
\end{equation}
where
\begin{eqnarray*}
	%	\matr{G}^* &=& \begin{bmatrix} \Gones & \matr{0} \\ \matr{0} & \Gtwos \end{bmatrix} \\ 
	\Gones &=& \begin{bmatrix} \Goneones & \matr{0} \\ \matr{0} & \Gonetwos \end{bmatrix}, \quad
	\Goneones = \begin{bmatrix} \matr{0} & \matr{0} \\ \matr{0} & \Gmoneone \end{bmatrix}, \quad 
	\Gonetwos = \begin{bmatrix} \matr{0} & \matr{0} \\ \matr{0} & \Gmonetwo \end{bmatrix}\\
	\Gtwos &=& \begin{bmatrix} \matr{0} & \matr{0} \\ \matr{0} & \Gmtwo \end{bmatrix}
\end{eqnarray*}
It follows that the reduced set of MME for the permute effects $\tbone$ are given by
\begin{equation}
	\Coneone \tbone = \Wone^\tp\Ptwo\matr{y} \label{eq:4}
\end{equation}
where $\Coneone = \Wone^\tp\Ptwo\Wone + \Gone^{*}, \Ptwo =
\matr{R}^{-1} - \matr{R}^{-1}\Wtwo(\Wtwo^\tp\matr{R}^{-1}\Wtwo+
\Gtwo^{*})^{-}\Wtwo^\tp\matr{R}^{-1}$ and
$(\Wtwo^\tp\matr{R}^{-1}\Wtwo + \Gtwo^{*})^{-}$ is any particular generalised
inverse of $\Wtwo^\tp\matr{R}^{-1}\Wtwo + \Gtwo^{*}$. It can be shown that 
\begin{displaymath}
	\Ptwo = \Vtwo^{-1} - \Vtwo^{-1}\Xtwo(\Xtwo^\tp\Vtwo^{-1}\Xtwo)^{-}\Xtwo^\tp\Vtwo^{-1}
\end{displaymath}
where $\Vtwo = \Ztwo\Gtwo\Ztwo^\tp + \matr{R}$ and
$(\Xtwo^\tp\Vtwo^{-1}\Xtwo)^{-}$ is any particular generalised inverse of
$\Xtwo^\tp\Vtwo^{-1}\Xtwo$. The matrix $\Ptwo$ has rank $n -
\rank(\Xtwo)$ and is unique, and it is the Moore-Penrose inverse of $\matr{T} =
\Mtwo\Vtwo\Mtwo$, where $\Mtwo = \matr{I}_n -
\Xtwo(\Xtwo^\tp\Xtwo)^{-}\Xtwo^\tp$. That is $\matr{T} =
\Ptwo^{+}$.

\subsubsection{Prediction and optimality criteria}
\label{sssec:designCriteria}

The aim is to find an optimal or near optimal design with respect to a $l-$vector of estimable functions $\matr{\pi} = \matr{D}_1\bone$ where $\matr{D}_1 = [\matr{D}_{11}\; \matr{0}]$ is a known $l \times p$ matrix, since $\bone$ is partitioned into objective and linked effects and $\matr{D}_{11}$ is a known $l \times q$ matrix. The vector of estimable functions, $\matr{\pi}$, involves terms from $O$, the set of objective effects and can comprise fixed, random or both fixed and random effects. For known $\soneone, \sonetwo, \stwo$ and $\matr{\sigma_r}$ 
\begin{equation}
	\matr{D}_1(\bone - \tbone) \sim \text{N}(\matr{0},\; \matr{\Lambda}) \label{eq:5}
\end{equation}
where $\matr{\Lambda} = \matr{D}_1\Coneone^{-}\matr{D}_1^\tp$ and $\Coneone^{-}$ a particular generalised inverse of the coefficient matrix of Eqn~(\ref{eq:4}).

A commonly used optimality criterion is the {\newAcrit} ($\mathcal{A}$) as this is usually considered appropriate in circumstances where all treatments are of equal interest, such as early stage plant breeding trials \citep{martinDesignExperimentsSpatial1986}. \cite{buenofilhoBlockDesignsRandom2007} show that a Bayesian design criterion for selection experiments that minimises the risk of an incorrect selection is equivalent to {\newAcrit} computed from the prediction error variance matrix for the vector of random entry (treatment) effects, $\matr{\Lambda}$ in Eqn~(\ref{eq:5}). In this case it follows that
\begin{displaymath}
	\mathcal{A} = \sum_{i}^{}\sum_{j<i} \pev(\tilde{\pi}_i - \tilde{\pi}_j)/l/(l-1)
\end{displaymath}
where $\pev{}$ refers to the prediction error variance of its scalar (or matrix) argument. A convenient form for computation of $\mathcal{A}$ is given by
\begin{displaymath}
	\mathcal{A} = \frac{2}{l-1}(\Tr(\matr{\Lambda}) -\frac{1}{l}\matr{1}_{l}^\tp\matr{\Lambda}\matr{1}_{l}).
\end{displaymath}

In the most general case, the prediction error variance matrix $\matr{\Lambda}$ can be computed directly from augmenting the mixed model Eqns~(\ref{eq:mm}) with $\matr{D}_1$ \citep{gilmourEfficientComputingStrategy2004a}. A prediction design matrix pre-processor is not yet available {\odw} but applications presented in this paper can be implemented through various arguments to the {\odw} function.

\subsubsection{Finding optimal designs}
\label{sssec:search}

Finding an optimal design in experiments with categorical treatments is a problem in combinatorial optimisation where a permutation of the terms in $P$ to the experimental units that is optimal with respect to the pre-specified criterion is sought. \cite{baileyDesignComparativeExperiments2008} defines a design as a function $T$ from $\Omega$ to $\mathcal{T}$, where $\Omega$ is the whole set of plots (i.e.~experimental units) and $\mathcal{T}$ is the whole set of treatments. Thus plot $\omega$ is allocated $T(\omega)$. A design can be represented as a permutation vector $\matr{p}$ which orders the rows of $\Wone$, and this is achieved by pre-multiplying $\Wone$ by a permutation matrix $\matr{\mathcal{P}}$. Formally a design instance, or configuration, $\xi$, is $\matr{\mathcal{P}}\Wone$. Then it follows that if $\matr{\mathcal{P}}^{(1)} = S_e(\matr{\mathcal{P}})$ is a row perturbation of $\matr{\mathcal{P}}$, a new design is $\xi^{(1)} = \matr{\mathcal{P}}^{(1)}\Wone$, where $S_e$ is a perturbation function. The perturbation function which is implemented in {\odw} exchanges two rows of $\matr{\mathcal{P}}$ subject to any pre-specified swap constraints (see \cite{butlerOptimalDesignExperiments2013} for more details). 

Let $\mathcal{D}$ be the set of designs attainable through all possible (and permissible) permutations of the rows of $\matr{\mathcal{P}}$. Even for modest designs it is impossible to enumerate all solutions in $\mathcal{D}$, hence an efficient evaluation strategy directed by an optimisation algorithm is necessary. Briefly, given a suitable starting design and choice of optimality criterion, $\mathcal{A}$, the optimal design search algorithm is given by:
\begin{enumerate}
	\item A method to calculate $\mathcal{A}$ for a given $\matr{\mathcal{P}}$,
	\item An interchange policy to move to a neighbouring configuration in $\mathcal{D}$,
	\item A sampling strategy for $\mathcal{D}$ and acceptance policy for new
	configurations,
	\item A stopping rule to terminate the search.
\end{enumerate}
Items 2-4 above encapsulate the search algorithm in the optimisation process. The search algorithm implemented in {\odw} is tabu search as \cite{coombesReactiveTABUSearch2002} found its performance was superior to simulated annealing for the class of designs considered in this paper. 

\subsection{An algorithm to construct optimal or near optimal designs for SESEs}\label{ssec:algall}
In this section an algorithm for the construction of designs for SESEs using a model-based approach is presented. The paradigm of model-based design requires the specification of a statistical model. The model which is used here is a LMM. This model is sufficiently general for most problems, but it can be easily extended for designs not considered in this paper. 

\subsubsection{A linear mixed model for SESEs}\label{sssec:thislmm}
Let $\matr{y}$ denote the $n-$vector of data. The linear mixed model for $\matr{y}$ is written as
\begin{equation}
	\matr{y} = \matr{X\tau} + \matr{Z_gu_g} + \matr{Z_pu_p} + \matr{e} \label{eq:thislmm}
\end{equation}	
where $\matr{\tau}$ is an $r-$vector of fixed effects with associated design matrix $\matr{X}$; $\matr{u_g}$ is an $m-$vector of random genetic effects with associated design matrix $\matr{Z_g}$; $\matr{u_p}$ is a $b-$vector of random peripheral effects with associated design matrix $\matr{Z_p}$ and $\matr{e}$ is the $n-$vector of errors. The vector $\matr{u_p}$ is partitioned into $v$ component vectors each of length $b_j$ and $b=\sum_{j=1}^{v} b_j$ and $\Var(\matr{u_p}) = \matr{G_p} = \oplus_{j=1}^{v} \sigma^2_{p_j}\matr{I}_{b_j}$. The vector $\matr{\tau}$ would minimally include mean parameter. The vector $\matr{u_p}$ usually contains effects associated with the plot structure of the SESE.   

The random genetic effects, contained in $\matr{u_g}$ are the total genotype effects. Genetic relatedness is incorporated by partitioning $\matr{u_g}$ into additive and non-additive effects given by 
\begin{equation}
	\matr{u_g} = \matr{u_a} + \matr{u_e}\label{eq:thislmmug}
\end{equation}
where $\matr{u_a}$ and $\matr{u_e}$ are the vectors of additive and non-additive genetic effects respectively (see for example, \cite{oakeyJointModellingAdditive2007}). Further it is assumed that
\begin{equation}
	\Var \left(\begin{array}{c}
		\matr{u_a} \\ \matr{u_e}
	\end{array}\right) = \left[\begin{array}{cc}
		\sigma^2_a\matr{G_r} & \matr{0}\\ \matr{0} & \sigma^2_e\matr{I}_m
	\end{array}\right]\label{eq:thislmmvar}
\end{equation}
where $\sigma^2_a$ and $\sigma^2_e$ are variance parameters associated with the additive and non-additive effects respectively. The matrix $\matr{G_r}$ is a known ($m\times m$) matrix which can be either a numerator relationship matrix (NRM), denoted by $\matr{A}$, a genomic relationship matrix (GRM) denoted by $\matr{K}$ or a joint relationship matrix (JRM) denoted by $\matr{H}$ (see for example \cite{meyerEstimatesGeneticTrend2018}). Hence it follows that 
\begin{equation}
	\Var(\matr{u_g}) = \sigma^2_a\matr{G_r} + \sigma^2_e\matr{I}_m\label{eq:thislmmric}
\end{equation}

Lastly, the variance of $\matr{e}$ is assumed to be $\matr{R}$.

\subsubsection{Description of an algorithm for design of a SESE}\label{sssec:desalg}
\begin{description}
	\item[Stage 1: Gathering the experiment attributes] 
	%
	%
	% this is a new section to replace the old one
	%
	In this stage attributes of the experiment are determined following ideas proposed by \cite[][chapter 1]{baileyDesignComparativeExperiments2008}. This is a key component of the algorithm, and \cite{baileyDesignComparativeExperiments2008} also stresses the importance of consultation with the scientist (here the breeder) as being the first stage of designing a comparative experiment. Her context was most likely for a classical approach to design construction, however this process is even more important when using a model-based design approach. For example, model-based design requires a pre-specified model, referred to as the ``design-model'', and it is strongly recommended that the design-model matches, as closely as possible, the so-called ``analysis-model''. The analysis-model can only be constructed in close collaboration with the domain expert(s). 
	
	\cite{baileyDesignComparativeExperiments2008} lists ``the ideal and the reality'' which entails determining 
	\begin{inlineenum}
		\item the purpose of the experiment;
		\item replication - the number of times a treatment (i.e.~a genotype) is tested;
		\item blocking - meaningful ways of dividing up the set of experimental units (EUs);
		\item constraints - associated with: the cost of an EU; the availability of test materials (i.e.~quantity of seed for each genotype); the conduct of the experiment (i.e.~number of EUs which can be harvested in one day, number of machines available for processing samples); pre-determined conditions set out by the breeder such as higher replication for ChVs
		
		Lastly, the plot and treatment structures are determined using either the ideas found in \cite{baileyDesignComparativeExperiments2008} or using the so-called Design Tableau approach of \cite{smithDesignTableauAid2018}. \cite{baileyDesignComparativeExperiments2008} defines the treatment and plot structure as meaningful ways of dividing up $\mathcal{T}$ and $\Omega$ respectively. The treatment structure for many SESEs is, in the notation of \cite{wilkinsonSymbolicDescriptionFactorial1973}, \texttt{U/Genotype}, where $U$ is the universal factor, defined by $U(\omega) = 1 $ for all $\omega$ in $\Omega$ \citep{baileyDesignComparativeExperiments2008}. The plot structure depends on the blocking structure and pre-determined constraints. Terms in the design and analysis models are usually based on the treatment and plot structures, though additional terms or more complex variance models can be added (see \cite{brienFormulatingMixedModels2009a}).   
	\end{inlineenum}
	
	\item[Stage 2: Allocating replication to genotypes] 
	%
	%
	% this is a new section to replace the old one
	%
	
	Allocation of replication to genotypes uses a LMM for the $\matr{\hat{u}_g}$ being the $m-$vector of BLUEs for the fixed mean parameters for each genotype. These are obtained from the fit of a working LMM for an $n-$vector of data, $\matr{y}$. This fit is referred to as step 2.1 and the LMM is simplistic but importantly it requires a valid, initial allocation of replication to genotypes. That is, the allocation must conform with the constraints set out in stage 1. In practice, a random but valid, allocation can be used and terms in the working LMM for $\matr{y}$ are for an unstructured experiment \citep[see][chapter 2]{baileyDesignComparativeExperiments2008} with a simple treatment structure. Hence the working LMM for $\matr{y}$ can be written as 
	\begin{equation}
		\matr{y}|\matr{u_g} = \matr{Z_g}\matr{u_g} + \matr{e} \label{eq:blues}
	\end{equation}
	where $\matr{u_g}$ is the $m-$vector of genotype mean parameters with associated design matrix $\matr{Z_g}$ and $\matr{e}$ is the $n-$vector of errors. The overall mean parameter is omitted to ensure estimability of $\matr{u_g}$ and the variance of $\matr{e}$ is assumed to be $\matr{R}$.  It follows that 
	\begin{eqnarray}
		\matr{\hat{u}_g} &=& \matr{\Omega} \matr{Z_g}^\tp\matr{R}^{-1}\matr{y} \nonumber\\
		\Var(\matr{\hat{u}_g}) &=& \matr{\Omega}\label{eq:bluesdist}
	\end{eqnarray}
	where $\matr{\Omega} = (\matr{Z_g}^\tp\matr{R}^{-1}\matr{Z_g})^{-1}$. The matrix $\matr{R}$ is usually assumed to be equal to $\sigma^2\matr{I}_n$, and  without loss of generality, $\sigma^2$ can be set to one. In this case $\matr{\Omega}$ is a diagonal matrix with elements $\omega_j, j=1,\ldots,m$, each being the inverse of an element of the so-called replication set for the experiment, denoted by $R_c = \{r_1,\ldots,r_c\}$. Further let $m_j$ be the number of genotypes in the $jth, j=1\ldots c$ replication set, noting that $m=\sum_{1}^{c} m_j$.
	
	In step 2.2, an optimal or near-optimal allocation of replicates to genotypes can be obtained by considering a LMM for $\matr{\hat{u}_g}$ given by 
	\begin{equation}
		\matr{\hat{u}_g} = \matr{1}_m \mu + \matr{D_g}\matr{u_a} + \matr{D_g}\matr{u_e} + \matr{\eta} \label{eq:puF}
	\end{equation}
	where $\matr{u_a}$ and $\matr{u_e}$ are the $m$-vectors of additive and non-additive genetic effects with associated design matrix $\matr{D_g}$ and variance matrix given in Eqn~(\ref{eq:thislmmvar}). The $m$-vector $\matr{\eta}$ represents the estimation errors of $\matr{\hat{u}_g}$. If $\matr{\hat{u}_g}$ is ordered genotypes within replication levels, then it follows that 
	$ \Var(\matr{\eta}) = \oplus_{j=1}^c r_j^{-1}\matr{I}_{m_j}$.
	
	For computational reasons Eqn~(\ref{eq:puF}) is rewritten as 
	\begin{equation}
		\matr{\hat{u}_g} = \matr{1}_m \mu + \matr{D_g}\matr{u_a} +  \matr{\eta}^{*} \label{eq:puFcomp}
	\end{equation}
	where $\matr{\eta}^{*} = \matr{D_g}\matr{u_e} + \matr{\eta}$ and it follows that 
	\begin{eqnarray*}
		\Var(\matr{\eta}^{*}) &=& \sigma^2_e\matr{D_g}^\tp\matr{D_g} + \oplus_{j=1}^c r_j^{-1}\matr{I}_{m_j} \\
		&=& \oplus_{j=1}^c (\sigma^2_e+r_j^{-1})\matr{I}_{m_j}
	\end{eqnarray*}
	since $\matr{1}_m^\tp\matr{D_g} = \matr{1}_m$ then $\matr{D_g}^\tp\matr{D_g} = \matr{I}_m$ and this is invariant under any valid interchange between the rows of $\matr{D}_g$.  Values for the variance parameters $\sigma^2_a$ and $\sigma^2_e$ can be chosen by the user or set to defaults, noting that $\sigma^2$ can be rescaled as necessary, though in our experience using the default value of one seems to work in practice. 
	
	\item[Stage 3: Allocation of plots to genotypes] 
	The third and final stage of the algorithm considers construction of an optimal or near-optimal design which allocates plots to genotypes based on the LMM set out in Eqn~(\ref{eq:thislmm}). Importantly, the initial configuration for this stage must comply with the allocation of replication of genotypes obtained from stage 2. The matrix $\matr{R}$ is usually chosen from the class of separable (first-order) autoregressive processes noting that $\sigma^2\matr{I}_n$ belongs to this class as a special (null) model. 
	
	It is usually preferable to use at least two steps in this stage to ensure that the design function is constrained to produce binary designs with respect to factors linked to the conduct of the experiment. Common examples include management blocks in field experiments, sequential runs for experiments conducted in glasshouses and multiple operators or machines for experiments conducted in the laboratory. This idea has been used by \cite[][chapter 6]{johnCyclicComputerGenerated1995} to construct an efficient Latinised resolvable row-column design. They used a two-step approach in which they started with an efficient Latinised $\alpha$-design, then used the algorithm of \cite{nguyenAlgorithmConstructingOptimal1993} to interchange treatments between plots within columns within replicates to produce the final design. 
\end{description}

\subsection{An algorithm to construct optimal or near optimal designs for MESEs}\label{ssec:mses}
An algorithm for the construction of designs for MESEs using a model-based approach is now presented. The algorithm is similar to that presented in section~\ref{ssec:algall}, but the linear mixed models are quite different and there is at least one additional stage which allocates sites to genotypes, given the allocation of replication to genotypes.  

\subsubsection{Factor analytic linear mixed model}\label{sssec:falmm}
It is assumed that the MESE comprises $t$ environments each constitutes a single trial. Let $\matr{y}$ denote the $n-$vector of data combined across all environments, where $\matr{y} = (\matr{y}_1^\tp,\matr{y}_2^\tp,\ldots,\matr{y}_t^\tp)^\tp$, and $\matr{y}_j^\tp$ is the $n_j-$vector of data for the $j$th environment. Note that $n=\sum_{j=1}^{t} n_j$. The linear mixed model for $\matr{y}$ is written as
\begin{equation}
	\matr{y} = \matr{X\tau} + \matr{Z_gu_g} + \matr{Z_pu_p} + \matr{e} \label{eq:falmm}
\end{equation}	
where $\matr{\tau}$ is an $v-$vector of fixed effects with associated design matrix $\matr{X}$; $\matr{u_g}$ is an $mt-$vector of random genetic effects with associated design matrix $\matr{Z_g}$; $\matr{u_p}$ is a $b-$vector of random peripheral effects with associated design matrix $\matr{Z_p}$ and $\matr{e}=(\matr{e}_1^\tp,\matr{e}_2^\tp,\ldots,\matr{e}_t^\tp)^\tp$, and $\matr{e}_j^\tp$ is the $n_j-$vector of errors for the $j$th environment. It is assumed that $\matr{u_p}$ is partitioned into $q$ component vectors each of length $b_j$ and $b=\sum_{j=1}^{q} b_j$ and $\Var(\matr{u_p}) = \matr{G_p} = \oplus_{j=1}^{q} \sigma^2_{p_j}\matr{I}_{b_j}$. The vector $\matr{\tau}$ would minimally include mean parameters for each environment. The vector $\matr{u_p}$ usually contains the set of effects associated with the plot structure of the MESE.  

Genetic relatedness is incorporated into Eqn~(\ref{eq:falmm}) using a similar argument and assuming that
\begin{displaymath}
	\Var \left(\begin{array}{c}
		\matr{u_a} \\ \matr{u_e}
	\end{array}\right) = \left[\begin{array}{cc}
		\matr{G_a}\otimes \matr{G_r} & \matr{0}\\ \matr{0} & \matr{G_e}\otimes \matr{I}_m
	\end{array}\right]
\end{displaymath}
where $\matr{G_a}$ and $\matr{G_e}$ are symmetric positive definite matrices for the between environment additive and non-additive GE effects respectively. Hence it follows that 
\begin{displaymath}
	\Var(\matr{u_g}) = \matr{G_a}\otimes \matr{G_r} + \matr{G_e}\otimes \matr{I}_m
\end{displaymath}

Lastly, the variance of $\matr{e}$ is assumed to be $\Var(\matr{e}) = \matr{R} = \oplus_{j=1}^t \matr{R}_j$, where $\matr{R}_j = \Var(\matr{e}_j)$. 

The most commonly used form for the between environment matrices for both additive and non-additive GE effects is the factor analytic (FA) model, which is given by $\matr{\Lambda_s}\matr{D_s}\matr{\Lambda_s}^\tp + \matr{\Psi_s}$ where $\matr{\Lambda_s}$ is the $t\times k_s$ matrix of environmental loadings, $\matr{D_s}$ is the diagonal matrix of factor score variances and $\matr{\Psi_s}$ is the diagonal matrix of specific variances for $s=a,e$. The FA variance model is based on a latent regression model for $\matr{u_s}$ given  by 
\begin{equation}
	\matr{u_s} = (\matr{\Lambda_s}\otimes \matr{I}_m)\matr{f_s} + \matr{\delta_s} \label{eq:16}
\end{equation}
where $\matr{f_s}$ and $\matr{\delta_s}$ are the $mk_s$- and $mt$-vectors of GE factor scores and specific GE effects for $s=a,e$ respectively and  
\begin{equation}
	\Var \left(\begin{array}{c}
		\matr{f_a} \\ \matr{\delta_a} \\ \matr{f_e} \\ \matr{\delta_e}
	\end{array}\right) = \left[\begin{array}{cccc}
		\matr{D_a}\otimes \matr{G_r} & \matr{0} & \matr{0} & \matr{0}\\ 
		\matr{0} &  \matr{\Psi_a}\otimes \matr{G_r} & \matr{0} & \matr{0}\\
		\matr{0} & \matr{0} & \matr{D_e}\otimes \matr{I}_m & \matr{0} \\
		\matr{0} & \matr{0} & \matr{0} & \matr{\Psi_e}\otimes \matr{I}_m
	\end{array}\right] \label{eq:16a}
\end{equation}
Fitting an FA model requires constraints which ensure identifiability of the model. A commonly used constraint sets $\matr{D_s}=\matr{I}_k$. For FA models of order $k_s>1$, additional constraints are imposed on the elements of $\matr{\Lambda_s}$. 

Despite the utility of FA models for modelling the variance of GE effects in MET datasets, their applicability for the model-based design of MESEs is limited, since in practice little is known a priori about the order of the FA models and the variance parameters associated with each FA model. Hence, a robust sub-model within the class of FA models which possess key features of an FA model and is commensurate with the aim of early stage selection experiments is used. This sub-model is a compound symmetric (CS) model, which is an FA model of order 1 (FA1), with constraints on both $\matr{\Lambda}$ and $\matr{\Psi}$. The CS model often referred to as the main effects and interaction model, is obtained by setting $k_s=1$, $\matr{\Lambda_s} = \matr{1}_t, \matr{D_s}=d_s$ and $\matr{\Psi_s} = \psi_s\matr{I}_t$ for $s=a,e$. It follows for $s=a,e$, the CS model has $\matr{G_s}=d_s\matr{J}_t + \psi_s\matr{I}_t$ where $\matr{J}_t = \matr{1}_t\matr{1}_t^\tp$ and $\matr{1}_t$ is the $t$-vector whose elements are one.      

For the CS model, Eqns~(\ref{eq:16}) and~(\ref{eq:16a}) are given by
\begin{equation}
	\matr{u_s} = (\matr{1}_t\otimes \matr{I}_m)\matr{f_s} + \matr{\delta_s} \label{eq:16b}
\end{equation}
and 
\begin{equation}
	\Var \left(\begin{array}{c}
		\matr{f_a} \\ \matr{\delta_a} \\ \matr{f_e} \\ \matr{\delta_e}
	\end{array}\right) = \left[\begin{array}{cccc}
		d_a\matr{G_r} & \matr{0} & \matr{0} & \matr{0}\\ 
		\matr{0} &  \psi_a\matr{I}_t\otimes \matr{G_r} & \matr{0} & \matr{0}\\
		\matr{0} & \matr{0} & d_e \matr{I}_m & \matr{0} \\
		\matr{0} & \matr{0} & \matr{0} & \psi_e\matr{I}_t\otimes \matr{I}_m
	\end{array}\right] \label{eq:16c}
\end{equation}
Substitution of Eqn~(\ref{eq:16b}) into Eqn~(\ref{eq:falmm}) gives

\begin{eqnarray}
	\matr{y} &=& \matr{X\tau} + \matr{Z_g}(\matr{1}_t\otimes \matr{I}_m)\matr{f_a} + \matr{Z_g\delta_a} + \matr{Z_g}(\matr{1}_t\otimes \matr{I}_m)\matr{f_e} + \matr{Z_g\delta_e} + \matr{Z_pu_p} + \matr{e} \nonumber \\
	&=& \matr{X\tau} + \matr{Z_f}\matr{f_a} + \matr{Z_g\delta_a} + \matr{Z_f}\matr{f_e} + \matr{Z_g\delta_e} + \matr{Z_pu_p} + \matr{e} \label{eq:17add} \\
	&=& \matr{X\tau} + \matr{Z_f}\matr{f_g} + \matr{Z_g\delta_g} + \matr{Z_pu_p} + \matr{e} \label{eq:17tot}	
\end{eqnarray}	
where $\matr{Z_f} = \matr{Z_g}(\matr{1}_t\otimes \matr{I}_m), \ \matr{f_g} = \matr{f_a}+\matr{f_e}$ and $\matr{\delta_g} = \matr{\delta_a} + \matr{\delta_e}$. It follows that  $\Var \left( \matr{f_g} \right) = d_a\matr{G_r} + d_e \matr{I}_m$ and $\Var \left( \matr{\delta_g} \right) = (\psi_a + \psi_e )\matr{I}_{mt}$.

Eqns~(\ref{eq:17add}) and~(\ref{eq:17tot}) provide two equivalent LMMs which are used for design construction. The former is used when the objective set includes additive genetic effects, while the latter is a computationally efficient when the objective set includes both additive and non-additive genetic effects. The efficiency comes from reducing the size of $\Coneone$ by $m$.   

Similarly robust variance models which are sub-models of those used for the analysis of MET datasets, are used for the variance models associated with $\matr{u_p}$ and $\matr{e}$. For example, \cite{smithPlantVarietySelection2021} uses variance heterogeneity linked to each environment for the components of $\matr{e}$ and $\matr{u_p}$, however, in the absence of prior information homogeneous variance models across environments are used. Spatial dependence for the errors can be used   \citep{cullisDesignEarlyGeneration2006,cullisDesignEarlyStagePlant2020}), however it may be preferable to use IID models for the errors (see \cite{cullisDesignEarlyStagePlant2020}).

\subsubsection{Description of an algorithm for design of a MESE}\label{sssec:mesealgdesc}
\begin{description}
	\item[Stage 1: Gathering the experiment attributes] 
	%
	%
	% this is a new section to replace the old one
	%
	This stage is very similar to stage 1 of the algorithm to design a SESE. There are two important differences:\begin{inlineenum}
		\item formulating the plot and treatment structures of the MESE ensuring that so-called anatomical variables \citep{smithDesignTableauAid2018} are accommodated; 
		\item meticulous definition and construction of factor(s) which define legal treatment interchanges during the design search. Rows of $\Wone$ for the permute set term(s) columns are only interchanged within levels of the so-called \texttt{swap} factor(s). 
	\end{inlineenum}
	
	The simplest plot structure for a MESE with no blocking factors within trials is \texttt{U/Site/Plot}, where $U$ is the universal factor, \texttt{Site} is a factor with $t$ levels and \texttt{Plot} is a factor which indexes the EUs within trials. The treatment structure is usually \texttt{Site/Genotype}, where \texttt{Genotype} is a factor with $m$ levels. This is commensurate with the FALMM presented in Eqn~(\ref{eq:falmm}), if $\matr{\tau}$ contains the site mean parameters, and overall mean parameter. The term $\matr{u_g}$ is the vector of nested genotype effects within site. In this instance, \texttt{Site} is an anatomical factor, hence cannot be used in the plot structure  \citep{smithDesignTableauAid2018}. This conflict is simply resolved by creating another factor which is aliased with \texttt{Site}. This new factor, referred to as \texttt{Environment} is then used in the treatment structure of the MESE. 
	
	The majority of MESEs, particulary those used for stages S1 and S2, require a complex set of constraints governing the allocation of sites to genotypes. Constraints arise from seed supply issues, adaptation of genotypes to environments and adequate control of the number of plots each genotype is tested within each trial. Construction of the factors which match these constraints can be non-trivial.  
	
	\item[Stage 2: Allocating replication to genotypes] 
	%
	%
	% this is a new section to replace the old one
	%
	The philosophy of the approach for this stage mirrors that used in stage 2 for the design of SESEs, where allocating of replication to genotypes uses an interim LMM for the $m-$vector of BLUEs for the fixed mean parameters of genotypes. However, in contrast to stage 2 for the SESE algorithm, a working and valid allocation of sites to genotypes is required to determine the variance of $\matr{\eta}$, being the vector representing estimation errors for $\matr{\hat{f}_g}$. This allocation is obtained using the model described in stage 3, but importantly, uses the initial allocation of replication to genotypes. 
	
	Given the working allocation of sites to genotypes, a LMM for the $n$-vector of data $\matr{y}$ is 
	\begin{equation}
		\matr{y}|\matr{f_g} = \matr{X\tau} + \matr{Z_ff_g} + \matr{Z_g\delta_g} + \matr{e} \label{eq:21} 
	\end{equation} 
	where $\matr{\tau}$ and $\matr{f_g}$ are vectors of fixed effects, containing the overall mean, mean parameters for sites and mean parameters for genotypes. The vector $\matr{\delta_g}$ contains random site by genotype interaction effects, where it is assumed that $\Var\left( \matr{\delta_g}\right) = d_g\matr{I_mt}$. Lastly $\matr{R}$ is assumed to be $\sigma^2\matr{I}_n$. Hence the MMEs for Eqn~(\ref{eq:21}) can be written as  
	\item[Stage 3: Allocation of sites to genotypes] 
	\item[Stage 4: Allocation of plots within sites to genotypes] 
\end{description}

\section{Results}\label{sec:results}
\subsection{SESE Illustrative example: Ascochyta blight disease screening nursery}\label{ssec:ascodesign}
This example is an experiment which investigates the tolerance to Ascochyta blight for the  set of test lines which entered GP3 in 2024 for the Chickpea Breeding Australian program. The main objective is to identify and select superior test lines for progression through this phase, however, a secondary objective is to use this experiment in the development of a training population to further enhance the accuracy of GS in the program. The experiment involves a high-throughout approach and will be conducted in a controlled environment facility at the Tamworth Agricultural Institute. Chickpea seedlings are placed in so-called plant rows (\code{PR}) which reside in crates (\code{C}). Each crate can fit either 18 or 21 PRs, and contains a medium which has been inoculated with \emph{Ascochyta rabei}. Crates are arranged on tables placed on tables (\code{T}) in a rectangular array which has two crate rows (\code{CR}) by 10 or 9 crate columns (\code{CC}. A total of 34 tables were placed in a rectangular array with 17 table rows (\code{TR}) and 2 table columns (\code{TC}). Tables within each TC were spaced apart to allow access during the course of the experiment and the two adjacent TCs were set apart to facilitate watering, measurement and routine inspection of seedlings during the experiment. To manage the workload, it was decided that the experiment be conducted across three sequential runs (\code{R}). The first five TRs for each TC were assigned to run 1, the next six to run 2 and the final 6 to run 3. To control heterogeneity within each run an additional blocking factor was imposed, namely zones (\code{Z}). Zones 1, 3 and 5 comprised tables in TC 1 for runs 1, 2 and 3 respectively, while zones  2, 4 and 6 comprised tables in TC 2 for runs 1, 2 and 3 respectively. Table~\ref{tab:asco.att} presents a summary of the plot factors and the total number of plots (EUs) for each run. Individual chickpea seedlings are assessed for disease symptoms after about 6 weeks. 
% latex table generated in R 4.4.1 by xtable 1.8-4 package
% Thu Nov 14 21:06:17 2024
\begin{table}[ht]
	\centering
	\scalebox{0.9}{
		\begin{tabular}{ccccccc}
			\toprule   & \multicolumn{6}{c}{Run} \\
			& \multicolumn{2}{c}{1} & \multicolumn{2}{c}{2} & \multicolumn{2}{c}{3}\\
			Attribute & Zone 1 & Zone 2 & Zone 3 & Zone 4 & Zone 5 & Zone 6 \\
			\midrule Plots (P) & 600 & 540 & 720 & 648 & 720 & 648 \\ 
			Tables (T) & 5 & 5 & 6 & 6 & 6 & 6 \\ 
			Crate Rows (CR) in T in Z & 2 & 2 & 2 & 2 & 2 & 2 \\ 
			Crate Columns (CC) in T in Z & 10 & 9 & 10 & 9 & 10 & 9 \\ 
			Plant Rows (PR) in Crates (C) in T in Z & 6 & 6 & 6 & 6 & 6 & 6 \\ 
			\bottomrule \end{tabular}
	}
	\caption{Experiment attributes table for the illustrative example.} 
	\label{tab:asco.att}
\end{table}

A total of 2528 chickpea TLs were available each with sufficient seed for two tests in the experiment. Four ChVs were also included in the experiment, and the breeder required there were six replicates in the experiment for each ChV. Ancestral information was available on all genotypes with 3151 individual genotypes in the pedigree. The total number of plots across the three runs was 3576, hence only 1324 TLs could be replicated twice, with the remainder having only one replicate. An optimal or near-optimal allocation of replication to TLs was obtained using the approaches outlined in stages 1 and 2 of the algorithm, subject to the constraint that each ChV had four replicates. 

The code snippet below provides the \odw\ call to find an optimal allocation of replication to genotypes. A more complete and annotated set of {\sf R} commands are provided in the supplementary material. Briefly the key arguments are  
\begin{inlineenum}
	\item \code{permute} set is resolved in the data frame factor called \code{name} and the NRM resides in the object \code{inVA} being a sparse form of $\matr{A}^{-1}$;
	\item \code{residual} specifies the variance model for the composite error term $\matr{\eta}^{*}$, noting the records must be sorted by the factor \code{repF} having three levels (1,2 and 6);
	\item \code{swap} specifies which factor(s) define legal exchanges of the \code{permute} set during the optimisation process, here a factor with two levels (\code{swp});
	\item \code{search} in this case a robust TABU search with a random walk as the local search strategy. \code{maxit} specifies the number of TABU loops;
	\item \code{R.param} and \code{G.param} provide values for $(\sigma_a^2, \sigma^2_e + \sigma^2/1 ,\sigma^2_e + \sigma^2/2 ,\sigma^2_e + \sigma^2/6 ) = (0.434, 1.2, 0.7, 0.367)$;
	\item \code{reorder} specifies columns in the initial data-frame that are to be permuted (at the termination of the search) in design order, parallel to the \code{permute} set;
	\item \code{data} provides the initial design as a data-frame in which to resolve the terms in the model formulae.
	\item \code{maxit} sets the number of TABU loops to 15
\end{inlineenum}

\begin{Code}
	odw(fixed = ~ 1, random = ~ vm(name, invA), 
	residual = ~dsum(~units|repF), 
	permute = ~ vm(name, invA), swap = ~swp, 
	reorder = c('type','stage'), R.param = sv, G.param = sv, 
	search = 'tabu+rw',maxit = 15, data = init.df)
\end{Code}
The cross tabulation of the factor (\code{swp}) controlling legal interchanges by the variance heterogeneity factor (\code{repF}) is shown below. 
% latex table generated in R 4.4.1 by xtable 1.8-4 package
% Wed Nov 13 14:27:04 2024
\begin{table}[ht]
	\centering
	\begin{tabular}{rrrr}
		\hline
		& \multicolumn{3}{c}{repF} \\
		swp & 1 & 2 & 6 \\
		\hline
		ChV &   0 &   0 &   4 \\ 
		TL & 1204 & 1324 &   0 \\ 
		\hline
	\end{tabular}
\end{table}
The cross tabulation of the final (\odw) and initial (random) allocation of replication to genotypes is presented below. Note the substantial number of re-allocations (1141 between TLs), suggesting that the initial configuration was quite sub-optimal. We revisit this proposition later in this section. 
% latex table generated in R 4.4.1 by xtable 1.8-4 package
% Thu Nov 14 15:44:52 2024
\begin{table}[ht]
	\centering
	\begin{tabular}{rrrr}
		\hline
		& \multicolumn{3}{c}{repF: initial} \\
		repF: final & 1 & 2 & 6 \\
		\hline
		1 &  63 & 1141 &   0 \\ 
		2 & 1141 & 183 &   0 \\ 
		6 &   0 &   0 &   4 \\ 
		\hline
	\end{tabular}
\end{table}
The breeder requested a plot of each ChV was in each zone and that no TL occurred more than once in each run. This meant that the design was an incomplete block design which is resolved for ChVs and zone. The breeder also stipulated the design should be optimal for the selection on total genetic effects. The set of plot factors for the experiment is \{\code{Run, Zone, Table, Crate, Crate Rows, Crate Columns, Plant Rows}\}. The working plot structure is given by, 
\begin{Code}
	R/Z/T/CR*CC/PR = R + R:Z + R:Z:T + R:Z:T:CR + R:Z:T:CC + 
	R:Z:T:CR:CC + R:Z:T:CR:CC:PR
	= R + Z + Z:T + Z:T:CR + Z:T:CC + 
	Z:T:C + Z:T:C:PR
\end{Code}
noting \code{Zone} is coded 1 to 6 and hence \code{R:Z} $\equiv$ \code{Z}. Previous experiments conducted in the same facility in 2022 and 2023 indicated that there was little variation between crate rows within tables within zones. However, post-blocking suggested  systematic effects between crate columns within zones were present in both experiments. The breeder confirmed these were visibly clear and provided additional evidence why these could have occurred. Therefore, the random model formulae was modified to,  
\begin{Code}
	R + Z + Z:T + Z:CC + Z:T:C 
\end{Code}
noting \code{Z:T:C:PR} $\equiv$ \code{units} and hence it has been removed from the random model formula. 

The first of two \odw\ calls which construct the design function with the desired properties is presented in the code snippet below. 
\begin{Code}
	odw(fixed=~ 1, random=~ ric(name, s1.ainv)  + Run,
	residual = ~units, permute=~ ric(name, s1.ainv), 
	swap = ~swp, R.param = sv, G.param = sv, 
	search = 'tabu+rw', maxit=10, data=step31.init.df)
\end{Code}
This call is referred to step 3.1. For this step, the initial configuration is provided in the \code{data} argument  is constructed ensuring that ChVs are resolved with respect to zones and replication of TLs is commensurate with the allocation from stage 2. 

Terms in the \code{random} argument include \code{name}, which is the \code{permute} factor and \code{Run}. The {\sf R} package \odw\ uses a functional model syntax which allows specification of the variance model for any term in the random and residual model formulae. The variance model for \code{name} is 
\begin{displaymath}
	\Var(\matr{u_g}) = \sigma_a^2\matr{A} + \sigma^2_e\matr{I}_{2532}
\end{displaymath}
using the \code{ric} variance function. The variance model for $\matr{e}$ is $\sigma^2\matr{I}_n$, where \code{units} is a special factor being the equality factor \citep[p171]{baileyDesignComparativeExperiments2008} and equal to \code{factor(seq(1, 3876))}. This is automatically generated by \odw\ and is a reserved factor name. The arguments \code{R.param} and \code{G.param} provide values for the variance parameters for the model terms in \code{random} and \code{residual}, and these are provided in the \code{data.frame}, \code{sv}. The values were based on the analysis of experiments conducted in 2021 and 2022 and given by $(\sigma_a^2, \sigma^2_e, \sigma^2_{r}, \sigma^2)^\tp = (0.234, 0.039, 0.400, 0.874)^\tp$, where $\sigma^2_r$ is the variance component of the run effects. 

The aim of step 3.1 is to construct a resolved design for TLs with two replicates with respect to runs while keeping the design resolved for ChVs and zones. The latter is achieved by setting the \code{swap} argument to \code{swp} and ensuring the initial configuration is resolved for ChVs and zones so that:
% latex table generated in R 4.4.1 by xtable 1.8-4 package
% Sat Nov 16 18:29:12 2024
\begin{table}[!htbp]
	\centering
	\begin{tabular}{rrrrrrrr}
		\hline
		& \multicolumn{7}{c}{swp} \\
		Zone & 1 & 2 & 3 & 4 & 5 & 6 & 7 \\
		\hline
		1 &   4 &   0 &   0 &   0 &   0 &   0 & 596 \\ 
		2 &   0 &   4 &   0 &   0 &   0 &   0 & 536 \\ 
		3 &   0 &   0 &   4 &   0 &   0 &   0 & 716 \\ 
		4 &   0 &   0 &   0 &   4 &   0 &   0 & 644 \\ 
		5 &   0 &   0 &   0 &   0 &   4 &   0 & 716 \\ 
		6 &   0 &   0 &   0 &   0 &   0 &   4 & 644 \\ 
		\hline
	\end{tabular}
\end{table}

% latex table generated in R 4.4.1 by xtable 1.8-4 package
% Sat Nov 16 18:32:26 2024
\begin{table}[!htbp]
	\centering
	\begin{tabular}{rrrrrrr}
		\hline
		& \multicolumn{6}{c}{Zone} \\
		name & 1 & 2 & 3 & 4 & 5 & 6  \\
		\hline
		CICA1841 &   1 &   1 &   1 &   1 &   1 &   1 \\ 
		ICC03996 &   1 &   1 &   1 &   1 &   1 &   1 \\ 
		KYABRA &   1 &   1 &   1 &   1 &   1 &   1 \\ 
		PBADRUMMOND &   1 &   1 &   1 &   1 &   1 &   1 \\ 
		\hline
	\end{tabular}
\end{table}
The \odw\ call for step 3.2 is presented below.  
\begin{Code}
	odw(fixed=~ 1, random=~ ric(name, s1.ainv)  + Run +
	Zone + Zone:Table + Zone:CrateCol + Zone:Table:Crate,
	residual = ~units, permute=~ ric(name, s1.ainv), 
	swap = ~swprun, R.param = sv, G.param = sv, 
	search = 'tabu+rw', maxit=120, data=step32.init.df)
\end{Code}
The initial configuration is provided in the \code{data} argument and is the design constructed in step 3.1. Terms in the random model formula, excluding \code{name} are present in the final plot structure. The values for the variance parameters provided in  \code{sv} were 
\begin{displaymath}
	(\sigma_a^2, \sigma^2_e, \sigma^2_{r}, \sigma^2_{z}, \sigma^2_{zt}, 
	\sigma^2_{zcc}, \sigma^2_{ztc}, \sigma^2)^\tp = 
	(0.234, 0.039, 0.400, 0.044, 0.043, 0.100, 0.315, 0.874 )^\tp 
\end{displaymath}
where the subscripts for each variance component have an obvious link to each term of the random model formula. The factor \code{swprun} defines legal exchanges during the optimisation process, and the cross tabulation below indicates its relationship with zones, whereby levels 7, 8 and 9 of this factor allow plots allocated to TLs in zones 1 and 2, 3 and 4, and 5 and 6 to be exchanged respectively. 

% latex table generated in R 4.4.1 by xtable 1.8-4 package
% Sun Nov 17 03:14:29 2024
\begin{table}[ht]
	\centering
	\begin{tabular}{rrrrrrrrrr}
		\hline
		& \multicolumn{9}{c}{swprun} \\
		Zone & 1 & 2 & 3 & 4 & 5 & 6 & 7 & 8 & 9 \\
		\hline
		1 &   4 &   0 &   0 &   0 &   0 &   0 & 596 &   0 &   0 \\ 
		2 &   0 &   4 &   0 &   0 &   0 &   0 & 536 &   0 &   0 \\ 
		3 &   0 &   0 &   4 &   0 &   0 &   0 &   0 & 716 &   0 \\ 
		4 &   0 &   0 &   0 &   4 &   0 &   0 &   0 & 644 &   0 \\ 
		5 &   0 &   0 &   0 &   0 &   4 &   0 &   0 &   0 & 716 \\ 
		6 &   0 &   0 &   0 &   0 &   0 &   4 &   0 &   0 & 644 \\ 
		\hline
	\end{tabular}
\end{table}

To assess the impact of genetic relatedness on theoretical (design) efficiency, we undertook a small study of four design types from a $2\times 2$ factorial of $\pm$ genetic relatedness at each stage of the design process. \cite{chanDesignFieldExperiments1999} considered robustness of model-based designs. Briefly, they define the model-based efficiency of a design as follows: if $D'$ is the optimal (or near-optimal) design found under an assumed model and $D^*$ is the optimal (or near-optimal) design found under the true model, then the model-based efficiency $E$ of design $D'$ is calculated as the ratio of $\mathcal{A}$ of $D^*$ under the true model to the $\mathcal{A}$ of $D'$ under the true model. That is,
\begin{displaymath}
	E_{D'} = \mathcal{A}_{D^*}/\mathcal{A}_{D'}
\end{displaymath}
In the context of model-based designs which use a LMM as the pre-specified model, this measure, $E$, can then be used to assess the robustness of designs to changes in variance models and changes in variance parameters within the class of LMMs. If $(AA, AI, IA, II)$ represent the four design settings, where 
\begin{inlineenum} 
	\item AA - used \odw\ allocation of replication to genotypes and  $\Var(\matr{u_g}) = \sigma_a^2\matr{A} + \sigma^2_e\matr{I}_{2532}$ in stage 2
	\item AI - used \odw\ allocation of replication to genotypes and  $\Var(\matr{u_g}) = \sigma^2_g\matr{I}_{2532}$ where $\sigma^2_g = \bar{a}\sigma^2_a + \sigma^2_e$ for $\bar{a} = \Tr(\matr{A})/2532$ in stage 2
	\item IA - used random allocation of replication to genotypes and  $\Var(\matr{u_g}) = \sigma_a^2\matr{A} + \sigma^2_e\matr{I}_{2532}$ in stage 2
	\item II - used random allocation of replication to genotypes and  $\Var(\matr{u_g}) = \sigma^2_g\matr{I}_{2532}$ where $\sigma^2_g = \bar{a}\sigma^2_a + \sigma^2_e$ for $\bar{a} = \Tr(\matr{A})/2532$ in stage 2
\end{inlineenum}
The results presented in Eqn~(\ref{Evals}) demonstrate the relative importance of using genetic relatedness in each stage of the design construction algorithm:  
\begin{equation}
	(E_{AA}, E_{AI}, E_{IA}, E_{II})^\tp = (1.000, 0.998, 0.966, 0.964)^\tp\label{Evals}
\end{equation}
For example, random allocation of replication to genotypes results in a substantial loss of model-based efficiency, whereas not using genetic relatedness to allocate plots to genotypes has a marginal influence on efficiency. The latter results are consistent with those of \cite{cullisDesignEarlyStagePlant2020} for designs with similar values for percent additive genetic variance, $p$-replication and accuracy.      

\section{Concluding remarks}\label{sec:concludingrem}

In this paper, we have presented a new algorithm for constructing optimal or near optimal designs for plant breeding selection experiments against a robust and sensible linear mixed model. This approach utilises information on genetic relatedness, as well as appropriately accommodates non-genetic sources of variation that may be evident across environments, thus reflecting the models used typically for analysis. The novelty of this approach lies within an additional step which accommodates efficient resource allocation of replication status to genotypes, which is undertaken prior to the allocation of plots to genotypes. 

The advantages of this approach have been exhibited using an illustrative example of a SESE, namely, an Ascochyta blight disease screening nursery, demonstrating that the informed allocation of replication status to genotypes will result in a substantial gain to model-based efficiency when compared to a random allocation of replication. Furthermore, there is a marginal influence on efficiency in the absence of information pertaining to genetic relatedness in the allocation of plots to genotypes. 

Current work has been done to show the advantages this method has with other categories of experiments, namely, MESEs and multiphase experiments, and we plan to update this paper with these examples and more in the near future.

\backmatter

\bmhead{Acknowledgements}

	We gratefully acknowledge David Tabah, Silvina Baraibar and David Moody, Intergrain for providing the motivating example. We also thank Kristy Hobson for providing the illustrative example for SESEs and Intergrain for part funding this work.

\bibliography{MyLibrary}% common bib file

\end{document}